# Diversifying Design of Nucleic Acid Aptamers Using Unsupervised Machine Learning

Siba Moussa[#,1], Michael Kilgour[#,1,$], Clara Jans[1,&], Alex Hernandez-Garcia[2], Miroslava Cuperlovic-Culf[3], Yoshua Bengio[2], and Lena Simine*[,1]

**Affiliations:**

[1]Department of Chemistry, McGill University, 801 Sherbrooke St. W, Montreal, Quebec, H3A 0B8, Canada.

[2]Montreal Institute for Learning Algorithms, 6666 St-Urbain, #200, Montreal, Quebec, H2S 3H1

[3]Digital Technologies Research Centre, National Research Council of Canada, 1200 Montreal Road, Ottawa, ON, K1A 0R6, Canada

Corresponding author: Lena Simine lena.simine@mcgill.ca

[#]authors contributed equally
[$]current affiliation: Department of Chemistry, New York University, New York City, New York, 10003, United States
[&]current affiliation: Department of Chemistry, University of Toronto, 80 Saint George St., Toronto, Ontario M5S 3H6, Canada

Abstract

Inverse design of short single-stranded RNA and DNA sequences (aptamers) is the task of finding sequences that satisfy a set of desired criteria. Relevant criteria may be, for example, the presence of specific folding motifs, binding to molecular ligands, sensing properties, etc. Most practical approaches to aptamer design identify a small set of promising candidate sequences using high-throughput experiments (e.g. SELEX), and then optimize performance by introducing only minor modifications to the empirically found candidates. Sequences that possess the desired properties but differ drastically in chemical composition will add diversity to the search space and facilitate the discovery of useful nucleic acid aptamers. Systematic diversification protocols are needed. Here we propose to use an unsupervised machine learning model known as the Potts model to discover new, useful sequences with controllable sequence diversity. We start by training a Potts model using the maximum entropy principle on a small set of empirically identified sequences unified by a common feature. To generate new candidate sequences with a controllable degree of diversity, we take advantage of the model's spectral feature: an 'energy' bandgap separating sequences that are similar to the training set from those that are distinct. By controlling the Potts energy range that is sampled, we generate sequences that are distinct from the training set yet still likely to have the encoded features. To demonstrate performance, we apply our approach to design diverse pools of sequences with specified secondary structure motifs in 30-mer RNA and DNA aptamers.

TOC Image

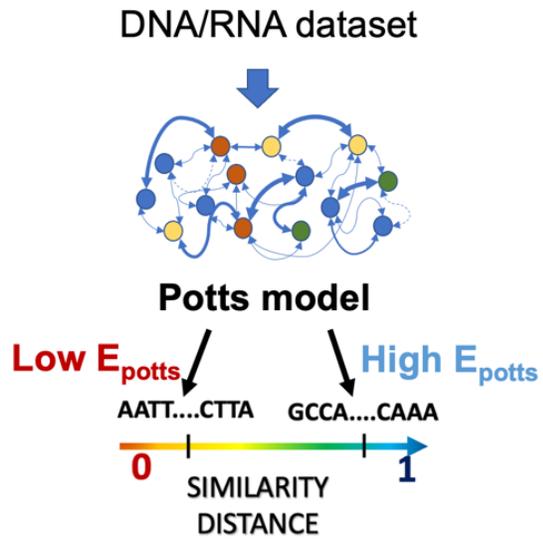

## Introduction

Biological polymer design strategies have been the topic of intense research for several decades [1-3] impacting many fields from pharmaceuticals to nanotechnology [4-9]. Single stranded nucleic acid oligomers with 20-100 residues, known as aptamers, are particularly useful for binding of ligands with high specificity and affinity. In therapeutics, aptamers are advantageous molecules due to their biocompatibility, accessibility and ease of modification [5,10]. Aptamers also offer high potential in diagnostics as they can serve as cheap and easily manufacturable molecular switches compared to their counterparts; enzymes and antibodies [11].

As is the case for many biomolecules, aptamer function is structure dependent. For aptamers, 3D folded configurations are determined not only by the sequence but also, to a large degree, by the surrounding conditions such as solvent type, ionic concentrations and composition, pH and temperature. Accurate structure prediction of nucleic acids is an actively developing field with many computational tools available to predict the secondary [12-15] and tertiary [16-22] structures, as well as to perform inverse (structure to sequence) design, for a review see Ref. 23. In practice, for applications in which aptamers bind a molecular target, sequences are selected using the Systematic Evolution of Ligands by Exponential Enrichment (SELEX) procedure which iteratively enriches DNA libraries with samples which preferentially bind the target molecular ligand [24,25]. SELEX data analysis is often strengthened through computational analysis of ligand docking, for a review see Ref. 26, combined with various flavors of machine learning inference techniques [27-29].

Indeed, new computational techniques for aptamer design rely increasingly on machine learning which offers computational advantages such as rapid search given appropriately trained models, ability to make nonlinear inferences, ability to be iteratively trained and improved. Iterative improvement is possible for example through active machine learning - a paradigm in machine learning in which the training dataset is iteratively expanded according to chosen or learned strategies [30-34]. Active learning is particularly valuable when data is expensive and difficult to obtain. Since the datasets generated by SELEX are often intended to be further expanded with fine-tuned and optimized sequences, the problem of design of nucleic acid aptamers is perfectly suited for active learning. To be effective, however, an active learning protocol requires a model that can learn from small amounts of data and direct experiments towards samples that are, at once, functionally promising, and distinct from those that are already in the training set.

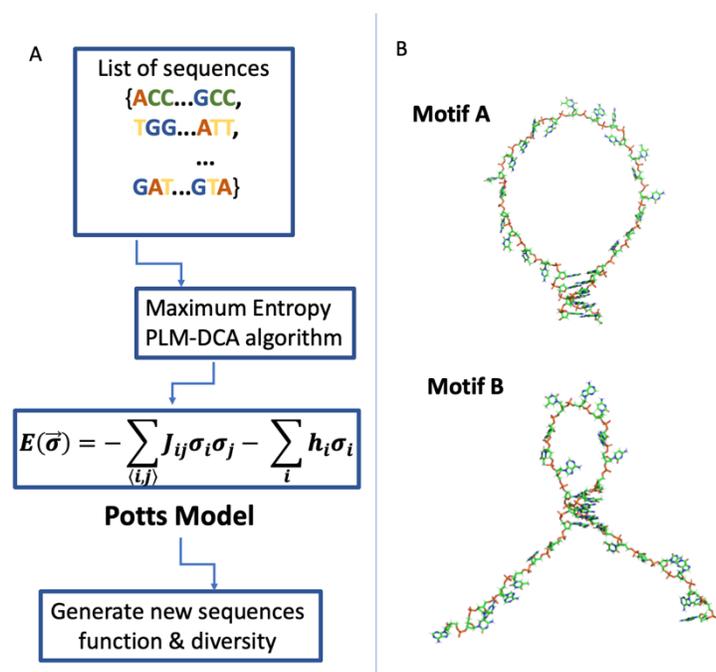

*Figure 1A. Flow diagram of the unsupervised learning approach employed in this work, the Potts model. B. 3D visualization of sample nucleic acid aptamers exhibiting the two contact motifs used in proof-of-concept experiments in this work.*

In this paper, we propose a computational approach to expanding small datasets of DNA or RNA aptamer sequences endowed with a specific property. To expand the dataset, we would like to search for new sequences following three rules: 1. New sequences must exhibit the desired feature with sufficiently high probability; 2. New sequences must be sufficiently different from those already included in the training set; 3. Only a limited number of attempts is allowed (due to high experimentation cost). To hypothesize new sequences with a controllable trade-off between diversity and accuracy, we repurpose an unsupervised learning modeling protocol that was used previously to predict secondary structure contacts in proteins [35]. The protocol is known to produce excellent results even when the training sets are very small (<1,000 data points), as is often the case with SELEX datasets. We demonstrate that in addition to capturing residue-residue interaction motifs related to function, our model can detect 'anomalies' in sequence space which will help us direct the search far from the training set. This model is called the Potts model and it emerges by maximizing the entropy subject to constraints supplied by the training dataset, for a schematic see Figure 1A.

To demonstrate that the Potts model offers a way to balance accuracy and diversity we explore secondary structure design of DNA and RNA 30-mer sequences for two motifs, see Figure 1B for an illustration. This paper is organized as follows: the Methods section presents the Potts model, the training protocol, and the technical details of sampling and analyses, Results and Discussion section presents our inverse RNA and DNA design experiments.

## Methods

### Unsupervised Learning: the Potts Model

The Potts Model may be thought of as a generalization of a fully connected recurrent neural network for n-state neurons or a spin-model. In our case, there are $n = 4$ states each corresponding to one of the bases: [A,T,G,C] for DNA and [A,U,G,C] for RNA and the number of 'spins' $N$ in the model corresponds to the number of residues in the sequences of interest. In Figure 1A we show the basic flow of this approach. The modeling is done by maximizing the entropy $S = -\sum_{\{\vec{\sigma}\}} P(\vec{\sigma}) \log P(\vec{\sigma})$ (where the sum runs over all possible sequences $\vec{\sigma} = \{\sigma_1, \sigma_2, \ldots, \sigma_N\}$ with $\sigma_i \in [1,2,3,4]$ ) subject to dataset-derived constraints. The constrains used in this work are the frequencies of appearance of bases $k$ at sequence position $i$: $f_i(k) = \frac{1}{M}\sum_{m=1}^{M} \delta(\sigma_i^m, k)$, and the frequencies of joint appearance of bases $(k, l)$ at pairs of sequence positions $(i, j)$: $f_{ij}(k, l) = \frac{1}{M}\sum_{m=1}^{M} \delta(\sigma_i^m, k)\delta(\sigma_j^m, l)$ where $\delta$ is the Kronecker delta that is equal to 1 when the two arguments are equal and equal to 0 otherwise, $M$ is the number of sequences in the training set, and $\sigma_i^m$ is the identity of i[th] residue in m[th] sequence $\vec{\sigma}$. In analogy with physical modeling, for a given sequence, we may define the Potts energy as

Equation 1 $\qquad E(\vec{\sigma}) = -\sum_{\langle i,j \rangle} J_{i,j}\, \sigma_i \sigma_j - \sum_{i=1}^{N} h_i \sigma_i$

where $\langle i,j \rangle$ runs over all pairs of indices $i$ and $j$, each taking integer values between 1 and N, and $\{J\}$ and $\{h\}$ are model parameters. The probability of sequence $\vec{\sigma}$ that maximizes the entropy is then given by

Equation 2
$$P(\vec{\sigma}) = e^{-E(\vec{\sigma})}/Z$$

where $Z = \sum_{\{\vec{\sigma}\}} e^{-E(\vec{\sigma})}$ is the partition function with the summation running over all sequences possible $\vec{\sigma}$. Various methods for Potts energy parameterization have been developed in the past [35-37], and most recently the Neural Potts Model [38]. In this work, we use the PLM-DCA pseudo-likelihood approach [39] in the Ising gauge.

**Direct Coupling Analysis**
To estimate the pairwise couplings between the bases, direct coupling analysis (DCA) is used. DCA ranks the base pairs according to their direct interaction strength, also known as direct information (DI). For each pair of residues $(i,j)$ the DI is calculated as follows:

Equation 3
$$DI_{ij} = \sum_{k,l=1}^{4} P_{ij}^{(dir)}(k,l) \ln \frac{P_{ij}^{(dir)}(k,l)}{f_i(k)f_j(l)}$$

with the direct joint probability distribution $P_{ij}^{(dir)}(k,l)$ defined as

Equation 4
$$P_{ij}^{(dir)}(k,l) = \frac{1}{Z_{ij}} \exp(-h_i(k) - h_j(l) - J_{ij}(k,l))$$

with the pseudo-partition function $Z_{ij} = \sum_{k,l=1}^{4} e^{-h_i(k)-h_j(l)-J_{ij}(k,l)}$ and

Equation 5
$$f_i(k) = \sum_{l=1}^{4} P_{ij}^{(dir)}(k,l)$$

Equation 6
$$f_j(l) = \sum_{k=1}^{4} P_{ij}^{(dir)}(k,l).$$

**Sampling: Markov Chain Monte Carlo (MCMC) Protocol with range restriction**
To obtain samples from specific regions of Potts model energy we have implemented a simple sampling scheme: a standard Metropolis-Hastings protocol (with 'temperature' parameter $\beta=0.5$) was restricted to sampling energies higher than a threshold energy. As we will show, varying this threshold amounts to turning a 'knob' for controlling accuracy of motif design and diversity of model predictions. For each value of energy threshold 50K MCMC steps were performed starting with a random seed sequence (results of these experiments showed no dependence on the choice of the initial seed), accepted sequences were ranked according to Potts energy, and only 100 lowest energy sequences were retained for analysis. The success rate of model prediction was quantified as the percentage of sequences (within the 100 lowest energy samples) that exhibited the correct secondary structure motif.

**Diversity Analysis: Cosine distance**
The cosine similarity is a measure commonly used to quantify similarity within biological sequences [40]. For two sequences $S_1$ and $S_2$ in a vector (one-hot) representation the cosine similarity measure $D_{cos}$ is given by

Equation 7 $$D_{cos} = 1 - \frac{S_1 \cdot S_2}{|S_1||S_2|}.$$

Here, we use the cosine similarity measure to evaluate the distance between sampled sequences and the reference (training) dataset. For each sequence, entries at all positions were one-hot encoded. An average of cosine distances to the reference set was recorded for each of the sampled sequences.

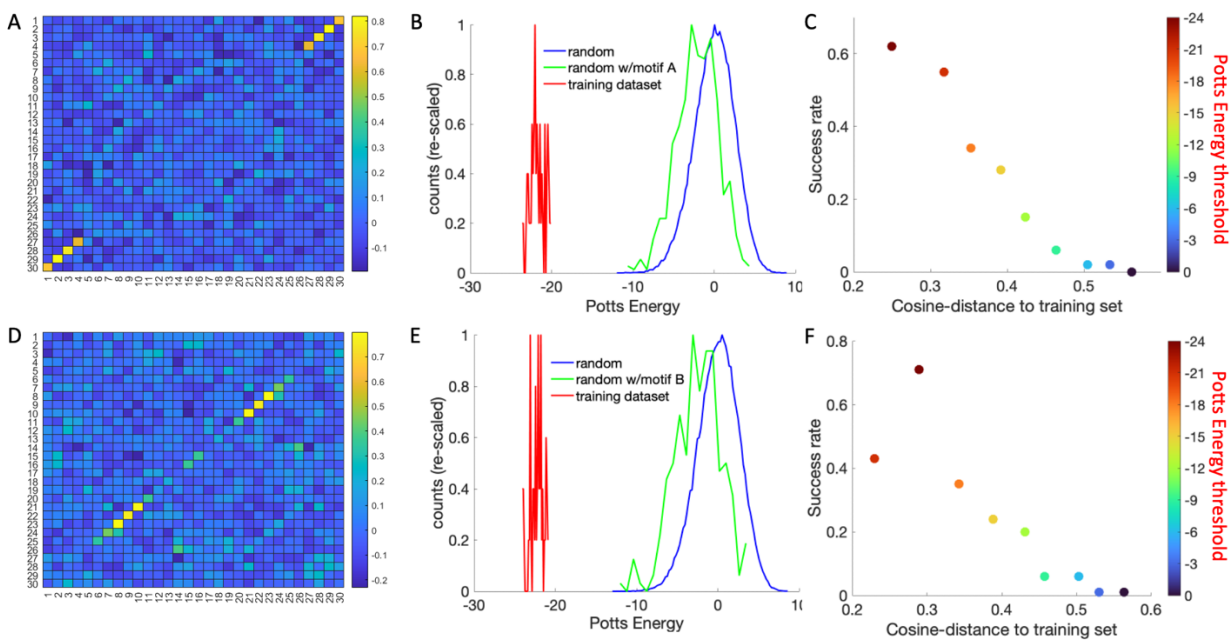

*Figure 2 RNA 30-mer aptamer inverse design experiments. Top row: results of experiments using Motif A dataset; Bottom row: results of experiments using Motif B dataset. Panels A and D: Direct Coupling Analysis (DCA): the axes show the residues indices (1-30) and the color bar indicates the direct information (DI) for two residues. Panels B and E: the energy structure of the Potts models: histogram traces (re-scaled by the value of the highest bar) for the sequences in the training set (red), randomly sampled sequences (blue), and the subset of random sequences which exhibits the correct folding motif (green). Panels C and F: Potts energy sampling experiments presented in 3D space of (y-axis) success rate of finding correct contact motif, (x-axis) cosine-distance to training set, (color bar) sampled Potts energy (threshold).*

**Experiments: Inverse RNA and DNA design task**

As proof of concept, we have chosen to work with mid-length 30-mer DNA and RNA aptamers and without loss of generality we have chosen to work with two folding motifs: Motif A with four residues on both termini in contact: 1-30, 2-29, 3-28, and 4-27, and Motif B with contacts between three pairs of residues: 8-23, 9-22, and 10-21, see Figure 1B for an illustration. We have constructed the training set by generating sequences randomly from a uniform distribution and retaining those that exhibited the required motif with no other acceptance criteria applied. The presence of a secondary structure motif was evaluated using RNAfold [41,42] for RNA and NUPACK [14] for DNA datasets. Imitating experimental datasets, the training sets were limited to only 50 samples.

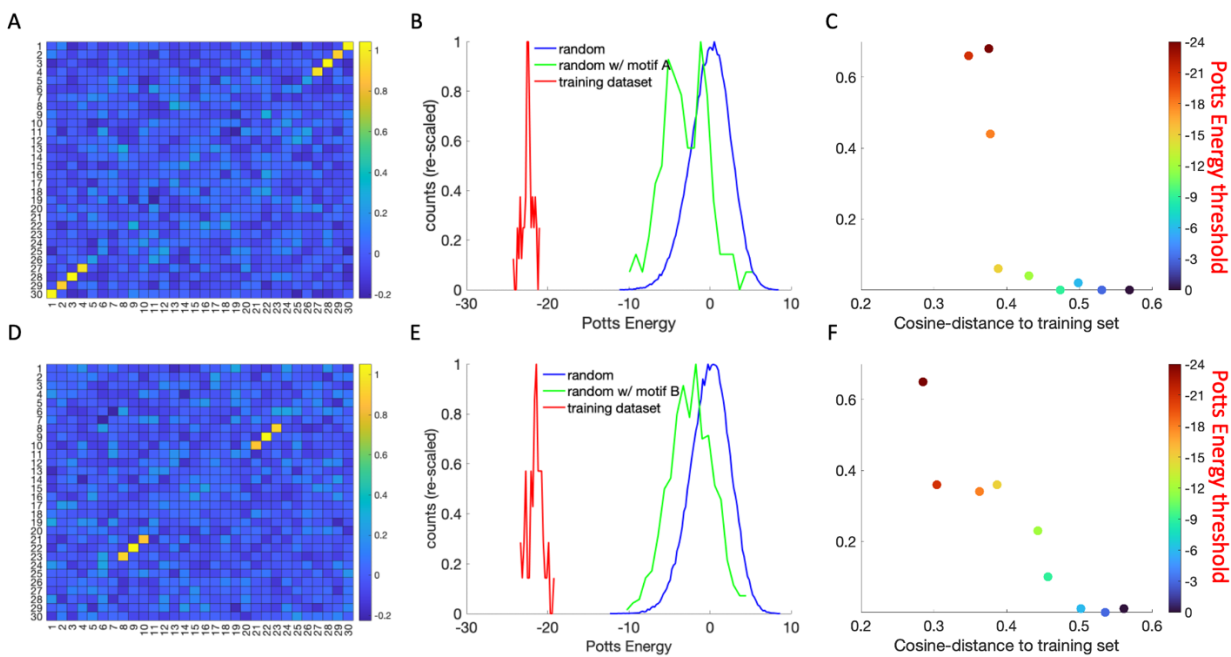

*Figure 3 DNA 30-mer aptamer inverse design experiments. Top row: results of experiments using Motif A dataset; Bottom row: results of experiments using Motif B dataset. Panels A and D: Direct Coupling Analysis (DCA): the axes show the residues indices (1-30) and the color bar indicates the direct information (DI) for two residues. Panels B and E: the energy structure of the Potts models: histogram traces (re-scaled by the value of the highest bar) for the sequences in the training set (red), randomly sampled sequences (blue), and the subset of random sequences which exhibits the correct folding motif (green). Panels C and F: Potts energy sampling experiments presented in 3D space of (y-axis) success rate of finding correct contact motif, (x-axis) cosine-distance to training set, (color bar) sampled Potts energy (threshold).*

**Results and Discussion**

Figures 2 and 3 summarize our results for the search for sequences exhibiting two secondary structure motifs (see Figure 1B for an illustration) in RNA and DNA 30-mers. In panels A and D the direct coupling analysis (DCA) for the training set is shown. The brightness of the matrix cells on this map indicates the strength of association between two residues (DI, see Equation 3) and indirectly it may be interpreted as a contact map. The matrix cells corresponding to motif A in the contact map are bright yellow in panels A and matrix cells corresponding to motif B light up in panels D of Figures 2 and 3, indicating that the Potts model has successfully inferred the secondary structure patterns from the training sets. In a close analogy with protein contact inference, here we focus on secondary structure in nucleic acid sequences. Nonetheless, we would like to emphasize that physical contact is only one example of possible information that can be encoded in the Potts model, the approach is general, and we expect it to work for datasets in which correlations between residues exist reflecting, e.g., binding to a target ligand.

Next, we examine the energy structure of our models. Panels B and E show the range of Potts energies for each model using the training set and a set of 100K sequences sampled randomly from a uniform distribution. In the figures, the red curve traces the histogram (re-scaled by the value of the highest bar) of the Potts model energies for the training dataset (50 samples). The blue curve traces the histogram (re-scaled by the highest bar) of energies sampled by a large uniformly distributed test set (100K samples). The green curve traces the re-scaled histogram of the subset of test-set sequences that happen to exhibit the relevant motif.

The first feature that stands out in panels B and E is the wide energy gap between the test set and the training set. The second feature is more subtle, but it is essential for our diversification protocol: the subset of random sequences that exhibit the encoded motif are shifted towards lower energies compared to the rest of the test set. These observations suggest that the Potts energy combines in a single scalar two bits of information: how distinct a sequence is from the training set, and how likely it is to exhibit the desired pattern – the lower the energy the higher the probability of finding a 'functional' sequence. These spectral properties of the Potts model suggest that by sampling a particular energy range of the Potts model it may be possible to balance the diversity of the generated samples (relative to the training set) with the probability of finding sequences containing the target motif.

In order to test the utility of Potts energy as a tuning parameter in our search for new functional and diverse sequences, we implement the standard Metropolis-Hastings Monte Carlo sampling protocol (see Methods) while restricting the accessible energy range from below and retaining only 100 lowest energy samples for analysis. The outcomes are shown in panels C and F of Figures 2 and 3 which show three dimensional plots which capture the characteristics of the sequences produced using our sampling protocol (see Methods for details). The three dimensions that we are exploring are: (y-axis) the success rate within the proposed set of sequences of matching the required folding motif, (x-axis) cosine-distance - a standard measure of how distinct the proposed sequences are from the training set (see Methods), and (color bar) the cut-off Potts energy used in the sampling protocol. The success rate estimate is based on the number of sampled sequences that exhibit the encoded motif within one hundred lowest energy sequences. Our motivation for limiting our analysis to only 100 lowest energy samples stems from the fact that in practice only a limited number of samples typically can be tested due to high experimental costs.

We make two key observations: 1. There is a monotonic inverse relationship between the success rate and the diversity of the samples indicating that although there is a trade-off between the two it can be controlled by varying the Potts energy cut-off value in the sampler; 2. There is a qualitative agreement between a standard measure of distance between sequences, cosine-distance, and the Potts energy - this positions the Potts energy as a sufficiently good witness of diversity of sequences relative to the dataset it was trained on. In practice, sampling higher energy regions of Potts energy will generate more diverse candidates at a price of lower probability of matching the desired property – this strategy is appropriate when test experiments are relatively cheap and simple to run and the value of finding a new distinct successful sequences is high; sampling the lowest energy region of the Potts model will lead to high probability of finding useful sequences but it will come at the expense of them carrying a lower degree of diversity - the suggested sequences will be likely similar to the training set; intermediate energy range offers a compromise between probability of successful prediction and exploration of the sequence space and it is most likely the useful regime for active learning. This concludes our presentation of a new computational approach based on unsupervised learning modeling to improving the search of chemical space of biopolymers with function and diversity in mind.

**Conclusions**

We have shown that the Potts model may be used to propose new nucleic acid sequences that are similar to a reference set in function but are distinct from it in sequence. We demonstrated this approach using examples inspired by RNA/DNA inverse design problem and hypothesized sequences with varied and controlled degree of success and diversity. Furthermore, we speculate that beyond secondary structure prediction, our approach may be applied to generic datasets in which the important information is captured by correlations between pairs of bases within the sequence, for example datasets which are constrained by binding affinity towards a particular molecular ligand. Being able to control the 'distance' of generated samples to the training set while maintaining sufficiently high success rate in identifying sequences with desired properties can help diversify the search process, lead it away from sequences that are already known, and facilitate sequence design tasks as well as the application of active learning to nuclei acid sequence design.

**Acknowledgments**

SB, LS and MCC acknowledge support from the National Research Council of Canada AI4Design program (CH-FY2021-AI4D-113 or NRC OCN-109-01), SB was in part supported through the Wares Science Innovation Prospectors Fund at McGill University. MK acknowledges support from NSERC PDF scholarship and CJ was supported through NSERC USRA scholarship. AHG and YB acknowledge support from CIFAR.